\documentclass[10pt, draft]{article}
\usepackage[T1]{fontenc}
\usepackage[latin1]{inputenc}
\usepackage{amsmath}
\usepackage{amsfonts}
\usepackage{amsthm}
\usepackage{epsfig}
\usepackage{graphicx, psfrag}
\usepackage{subfigure}
\usepackage{wrapfig}
\numberwithin{equation}{section}
\newtheorem{theorem}{Theorem}

\newtheorem{lemma}{Lemma}

\newtheorem{definition}{Definition}
\newtheorem{corollary}{Corollary}

\begin{document}
\title
\title{\center{\bf\large{A constraint variational problem arising\\ in stellar dynamics\\}}}
%\date
%\date{\center{March 5th, 2006}\\}
\author
\author{\center{Mahir Had\v zi\'c \\Division of Applied Mathematics, \\Brown University, Providence, 02912 RI USA\\}}
%\begin{center}
%\bf{Tentative}
%\end{center}
%\begin{center}
%Mahir Had\v zi\'c
%\end{center}
\begin{abstract}
We use the compactness result of A.~Burchard and Y.~Guo (cf.~\cite{BuGu}) to analyze the reduced 'energy' functional arising naturally in the stability analysis of steady states of the Vlasov-Poisson system (cf.~\cite{SaSo} and \cite{Ha}). We consider the associated variational problem and present a new proof that puts it in the general framework for tackling the variational problems of this type, given by Y.~Guo and G.~Rein (cf.~\cite{Re1} and \cite{Re2}).
\end{abstract}
\section{Introduction and statement of the result}
Our starting point is the Vlasov-Poisson system
\begin{equation}\label{vp1}
\partial_tf +v\cdot\nabla_xf-\nabla_xU\cdot\nabla_vf=0,
\end{equation}
\begin{equation}\label{vp2}
\Delta U=4\pi\rho,\quad \lim_{|x|\rightarrow\infty}U(t,x)=0,
\end{equation}
\begin{equation}\label{vp3}
\rho(t,x)=\int f(t,x,v)dv,
\end{equation}
where the dynamic variable $f=f(t,x,v)$ is the number density of a large ensemble of particles which interact by the gravitational potential $U=U(t,x)$. The variables $x$, $v\in\mathbb R^3$ denote position and velocity, $t\in\mathbb R$ is the time variable, and $\rho=\rho(t,x)$ is the spatial mass density induced by $f$.

Questions of nonlinear stability of stationary solutions to the Vlasov-Poisson system initiated many developments in recent years ( cf.~\cite{Re1} for a self-contained overview). The core idea was to recognise that a whole class of \textit{polytropic} steady states can be obtained as minimizers of so-called energy-Casimir functionals. Once this connection is established, one makes use of the minimization property of steady states to deduce their non-linear stability. 
We introduce the notation
\begin{equation*}
L_+^p(\mathbb R^n,M):=\big\{f\in L^p(\mathbb R^n); f\geq0\;\;\textrm{a.e.}, ||f||_{L^p(\mathbb R^n)}=M\big\},
\end{equation*}
and define kinetic and potential energy
\begin{equation*}
{\cal E}_{\textrm{kin}}(f):=\frac{1}{2}\int\int|v|^2f(x,v)\,dvdx,
\end{equation*}
\begin{equation*}
{\cal E}_{\textrm{pot}}(f):=-\frac{1}{8\pi}\int|\nabla U_f(x)|^2\,dx=-\frac{1}{2}\int\int{\frac{\rho_f(x)\rho_f(y)}{|x-y|}}\,dxdy,
\end{equation*}
where $\rho_f(x)=\int f(x,v)dv$. It is well known that the total energy
\begin{equation*}
{\cal E}(f):={\cal E}_{\textrm{kin}}(f)+{\cal E}_{\textrm{pot}}(f),
\end{equation*}
is conserved along the solutions of the Vlasov-Poisson system (\ref{vp1})--(\ref{vp3}). By abuse of notation we shall also write
\begin{equation*}
{\cal E}_{\textrm{pot}}(\rho)=-\frac{1}{2}\int\int{\frac{\rho(x)\rho(y)}{|x-y|}}\,dxdy.
\end{equation*}
The polytropic solutions are solutions of the following form
\begin{equation*}
f_{\mu}(x,v):=(E_0-|v|^2/2-U(|x|))_+^{\mu},
\end{equation*}
where $(f)_+$ denotes the positive part of the function $f$, $E_0\in\mathbb{R}$ is a constant and  $-1/2<\mu<7/2$. For a certain range of $\mu$ the polytropes with prescribed mass $M$ were shown to be minimizers of the energy-Casimir functional
\begin{equation*}
{\cal E}_C(f)={\cal E}(f)+ \int Q(f(x,v))\,dxdv
\end{equation*}
under the constraint $f\in L_+^{1}(\mathbb R^6,M)$. By formulating the problem in terms of spatial densities $\rho=\int f(.,v)dv$ in \cite{Re2}, the author naturally reduced it to the problem of minimizing a functional of the form
\begin{equation}\label{eq:reduciran}
{\cal E}_C^r(\rho)=\int\Phi(\rho(x))\,dx+{\cal E}_{\textrm{pot}}(\rho) 
\end{equation}
under the constraint  $\rho\in L_+^{1}(\mathbb R^3,M)$. The notion of \textit{reduction} and the exact relations between $Q$ and $\Phi$ are carefully analyzed in \cite{Re2}, where a concentration-compactness type argument is used to deal with the variational problem. A.~Burchard and Y.~Guo showed that it suffices to restrict the minimization procedure to the set of symmetrically decreasing functions $\rho$ (cf.~\cite[Thm.~1]{BuGu}). This makes the solution of the reduced variational problem simpler. In the review paper \cite{Re1} this technique is put in a formal framework involving several steps, indicating the possible genericity of this approach.
In \cite{SaSo}, \'O.~S\'anchez and J.~Soler approach the stability question by regarding the problem of minimizing the energy ${\cal E}(f)$ over the set of positive functions with prescribed $L^1$ and $L^{1+1/\mu}$ norms, with $\mu\in]0,7/2[$. More precisely, they minimize the functional ${\cal E}$ over the constraint set
\begin{equation*}
\Gamma_{M,J}^{\mu}:=L_+^1(\mathbb R^6,M)\cap L_+^{1+1/\mu}(\mathbb R^6,J).
\end{equation*}
We denote
\begin{equation}\label{eq:brat}
I_{M,J}^{\mu}:=\inf{\Big\{{\cal E}(f);\;\; f\in\Gamma_{M,J}^{\mu}\Big\}}.
\end{equation}
 The crucial difference to the method used by Y.~Guo and G.~Rein is that in this case we have to deal with two simultaneous constraints. The crux of the method is to reduce the energy functional to a functional defined only over spatial densities $\rho$ and at the same time to keep only one constraint in the minimization procedure. The new \textit{equivalent} problem, derived in \cite{SaSo}, is to minimize
\begin{equation*}
{\cal E}_{J}^{\mu}(\rho):=\frac{K_{1,1}}{J^{\frac{2(\mu+1)}{3}}}\Big(\int\rho^{\frac{2\mu+5}{2\mu+3}}\,dx\Big)^{\frac{2\mu+3}{3}} + {\cal E}_{\textrm{pot}}(\rho)
\end{equation*}
over the constraint set
\begin{equation*}
{\cal F}_{M}^{\mu}:=L_+^1(\mathbb R^3,M)\cap L^{\frac{2\mu+5}{2\mu+3}}(\mathbb R^3).
\end{equation*}
$K_{1,1}$ is just a constant arising from the reduction procedure and its value does not play a role for the rest of the paper. For details, see \cite{SaSo}.
We denote
\begin{equation}\label{eq:red}
R_{M,J}^{\mu}:=\inf\Big\{{\cal E}_{J}^{\mu}(\rho);\rho\in {\cal F}_{M}^{\mu}\Big\},
\end{equation}
and
\begin{equation*}
\Psi(\rho)=\int\rho^{\frac{2\mu+5}{2\mu+3}}\,dx,\quad\quad K:=\frac{K_{1,1}}{J^{\frac{2(\mu+1)}{3}}}.
\end{equation*} 
The above mentioned equivalence holds in the following sense:
\begin{lemma}[Equivalence of the variational principles]\label{co:kraj}
The variational problems of minimsing ${\cal E}$ over the constraint set $\Gamma_{M,J}^{\mu}$ and ${\cal E}_J^{\mu}$ over the constraint set ${\cal F}_M^{\mu}$, are equivalent in the following sense:
\begin{enumerate}
\item The infima $I_{M,J}^{\mu}$ and $R_{M,J}^{\mu}$ (cf. (\ref{eq:brat}) and (\ref{eq:red}) respectively) coincide, i.e. $I_{M,J}^{\mu}=R_{M,J}^{\mu}$.
\item If $(f_n(\cdot,\cdot))\subset\Gamma_{M,J}^{\mu}$ is a minimizing sequence of the functional ${\cal E}$ then the sequence $(\rho_n(\cdot))=(\int f_n(\cdot,v)dv)\subset {\cal F}_M^{\mu}$ is a minimizing sequence for the reduced functional ${\cal E}_J^{\mu}$. 
\item The functional ${\cal E}$ has a minimum over $\Gamma_{M,J}^{\mu}$ if and only if the functional ${\cal E}_J^{\mu}$ has a minimum over the constraint set ${\cal E}_M^{\mu}$. In that case the corresponding minimizers $f(\cdot,\cdot)$ and $\rho(\cdot)$ also verify $\rho(\cdot)=\int f(\cdot,v)dv$.
\end{enumerate}
\end{lemma} 
For a proof cf.~\cite{SaSo}. 
The aim of this paper is to show how the analysis of the reduced problem again fits into the general framework of the result of A.~Burchard and Y.~Guo.
Before stating the main theorem, we introduce the following definition:
\begin{definition}
Let $n\in\mathbb N$. A mapping $T$ is called \textit{translation} if there exists a shift vector $a\in\mathbb R^n$ such that $Tf(\cdot)=f(\cdot-a)$, for any function $f:\mathbb R^n\to\mathbb R$. 
\end{definition}
 We shall prove the following theorem:
\begin{theorem}\label{th:c}
Let $(\rho_n)\subset {\cal F}_M^{\mu}$ be a minimizing sequence of the functional ${\cal E}_J^{\mu}$ and let $\mu\in]0,7/2[$. Then there exists a sequence of translations $T_n$, a subsequence of $(\rho_n)$ (which we denote again by $(\rho_n)$), and $R>0$ such that 
\begin{equation*}
\int_{|x|\geq R} T_n\rho_n(x)\,dx\to0\quad\textrm{as}\quad n\to\infty, 
\end{equation*}
\begin{equation*}
T_n\rho_n\to\rho_0\quad\textrm{strongly}\quad in\quad L^{\frac{2\mu+5}{2\mu+3}}(\mathbb R^3),
\end{equation*}
and
\begin{equation*}
\int_{B_R}{\rho_0(x)}\,dx=M\quad\textrm{and}\quad\mathrm{supp}(\rho_0)\subset B_R.
\end{equation*}
In addition to this,
\begin{equation*}
\nabla U_{T_n\rho_n}\to\nabla U_{\rho_0}\quad\textrm{strongly in}\quad L^2(\mathbb R^3)\;\;\;\textrm{as}\quad n\to\infty,
\end{equation*}
and $\rho_0$ is a minimizer of the functional ${\cal E}_{J}^{\mu}$ over the set ${\cal F}_{M}^{\mu}$.
\end{theorem}

In \cite{SaSo} the authors used a concentration-compactness type argument in the spirit of \cite{Re2}, but here we give a different proof.
\section{Proof of the main result}

The crucial part of the proof is to carefully examine the behavior of the spherically symmetric minimizing sequences and then apply \cite[Thm.~1]{BuGu}. In order to emphasize the general nature of this method we follow the setup provided in \cite{Re1}, where the author analyzed the problem of minimizing (\ref{eq:reduciran}):

\subsubsection*{Step1: Concentration implies compactness}
The following lemma lemma will be used to treat the behavior of the potential energy along the spherically symmetric minimizing sequences.
\begin{lemma}\label{lm:novembar}
Fix any $0<n<5$ and let $(\rho_j)\subset L_+^{1+1/n}(\mathbb R^3,M)$ be a sequence of functions such that $\rho_j\rightharpoonup\rho_0$ weakly in $L^{1+1/n}(\mathbb R^3)$. Assume that
\begin{equation*}
\lim_{j\to\infty}\int_{|x|\geq R}\rho_j=0
\end{equation*}
for some $R>0$, i.e., the mass remains asymptotically concentrated in the ball of radius $R$. Then
\begin{equation*}
{\cal E}_{\textrm{pot}}(\rho_j-\rho_0)\to0,\quad j\to\infty.
\end{equation*}
\end{lemma}
\begin{proof}
Let us set $\sigma_j:=\rho_j-\rho_0$. For $\delta>0$ let us split the integral
\begin{equation*}
I_j:={\cal E}_{\textrm{pot}}(\sigma_j)=-\int\frac{\sigma_j(x)\sigma_j(y)}{|x-y|}\,dxdy
\end{equation*}
into three parts
\begin{equation*}
I_j=I_{j,1}+I_{j,2}+I_{j,3}
\end{equation*}
where
\begin{center}
$|x-y|<\delta$ for $I_{j,1}$,\quad $|x-y|\geq\delta\wedge(|x|\geq R\vee|y|\geq R)$ for $I_{j,2},$
\end{center}
\begin{center}
$|x-y|\geq\delta\wedge|x|\geq R\wedge|y|\geq R$ for $I_{j,3}$.
\end{center}
Obviously, $\int\sigma_j\,dx\leq2M$ for every $j$.
Since $2n/(n+1) + 2/(n+1)=2$, we get by Young's inequality
\begin{equation*}
|I_{j,1}|\leq C||\sigma_j||_{1+1/n}^2||\textbf{1}_{B_{\delta}}|.|^{-1}||_{n+1/2}\leq C\Big(\int_{0}^{\infty}r^{\frac{3-n}{2}}\,dr\Big)^{2/(n+1)}\to 0
\end{equation*}
if $\delta\to0$, uniformly in $j$ (note that $\textbf{1}_{A}$ stands for the characteristic function of the set $A$ and $B_{r}$ refers to the ball of radius $r$, for $r>0$). Furthermore,
\begin{equation*}
|I_{j,2}|\leq\frac{2M}{\delta}\int_{|x|\geq R}|\sigma_j(x)|\,dx\to0,
\end{equation*}
as $j\to\infty$, for any fixed $\delta$. Finally by Hölder's inequality
\begin{equation*}
|I_{j,3}|=\Big|\int\sigma_j(x)h_j(x)\Big|\leq||\sigma_j||_{L^{1+1/n}(\mathbb R^3)}||h||_{L^{1+n}(\mathbb R^3)},
\end{equation*}
where, in a pointwise sense
\begin{equation*}
h_j(x):=\textbf{1}_{B_R}(x)\int_{|x-y|\geq\delta}\textbf{1}_{B_R}(y)\frac{\sigma_j(y)}{|x-y|}dy\to0
\end{equation*}
which follows by the weak convergence of $\sigma_j$ and the fact that we are integrating $\sigma_j$ against a test function in $L^{1+n}$. But, since $h_j\leq\frac{2M}{\delta}$ for every $j$, we conclude by Lebesgue's dominated convergence theorem that $h_j\to0$ in $L^{1+n}$ and thus $|I_{j,3}|\to0$ as $j\to\infty$.
The lemma is proven.
\end{proof}

\subsubsection*{Step 2: Behavior under rescaling}

In analogy to \cite{Re1} (Section $5$, Step $3$) one needs to examine the behavior of the involved functional under  scaling. The statement and proof can be found in \cite{SaSo}. For the sake of completeness we state the result.
\begin{lemma}\label{lm:scaling}
The infima $I_{M,J}^{\mu}$ and $R_{M,J}^{\mu}$ verify:
\begin{enumerate}
\item\label{it:u} $I_{M,J}^{\mu}=R_{M,J}^{\mu}=M^{\frac{7-2\mu}{3}}J^{\frac{2(\mu+1)}{3}}I_{1,1}^{\mu}$,
\item\label{it:v} $-\infty<I_{M,J}^{\mu}=R_{M,J}^{\mu}<0$.
\end{enumerate}
\end{lemma}
\begin{corollary}\label{co:bitno}
Any minimizing sequence of ${\cal E}_{J}^{\mu}$ over ${\cal F}_{M}^{\mu}$ is uniformly bounded in $L^{\frac{2\mu+5}{2\mu+3}}(\mathbb R^3)$.
\end{corollary}
\begin{proof}
Let $(\rho_n)$ be a minimizing sequence. From the proof of Lemma~\ref{lm:scaling} (cf.~\cite{SaSo}) it is then easy to conclude that every minimizing sequence is uniformly bounded in $L^{\frac{6}{5}}(\mathbb R^3)$ and that $({\cal E}_{\textrm{pot}}(\rho_n))$ is also uniformly bounded. Finally, from the definition of ${\cal E}_{J}^{\mu}$ we deduce the claim. 
\end{proof}

\subsubsection*{Step 3: Spherically symmetric minimizing sequences remain concentrated}
Now we state the crucial concentration argument for spherically symmetric minimizing sequences of the reduced problem. 
\begin{lemma}\label{lm:sim}
Let us define
\begin{equation*}
R_0:=\frac{M^2}{-kR_{J,M}^{\mu}}\quad\textrm{where}\quad k:=7/3-2\mu/3.
\end{equation*}
Let $\rho\in {\cal F}_{M}^{\mu}$ be spherically symmetric, $R'>0$ and define
\begin{equation*}
m:=\int_{|x|\geq R'}\rho(x)\,dx.
\end{equation*}
Then the following inequality holds
\begin{equation*}
{\cal E}_{J}^{\mu}(\rho)\geq R_{J,M}^{\mu}+m(M-m)\big[\frac{1}{R_0}-\frac{1}{R'}\big].
\end{equation*}
If $R'>R_0$ then for every spherically symmetric minimizing sequence $(\rho_n)\subset {\cal F}_{M}^{\mu}$ of the functional ${\cal E}_{J}^{\mu}$ we have 
\begin{equation*}
\lim_{n\to\infty}\int_{|x|\geq R'}\rho_n(x)\,dx=0.
\end{equation*}
\end{lemma}
\begin{proof}
Although the statement of this lemma is completely analogous to the Step $4$, Section $5$ of \cite{Re1}, the proof is based on somewhat more complicated arguments, due to the more complicated nature of the scaling relations in Lemma \ref{lm:scaling}. 
We define $\rho_1:=\textbf{1}_{B_{R'}}\rho$ and $\rho_2:=\rho-\rho_1$, and also
\begin{equation*}
\alpha_1:=\frac{\int\rho_1^{\frac{2\mu+5}{2\mu+3}}\,dx}{\int\rho^{\frac{2\mu+5}{2\mu+3}}\,dx}\;\;\textrm{and}\;\;\alpha_2:=\frac{\int\rho_2^{\frac{2\mu+5}{2\mu+3}}\,dx}{\int\rho^{\frac{2\mu+5}{2\mu+3}}\,dx}.
\end{equation*}
By keeping in mind that
\begin{equation*}
K=\frac{K_{1,1}}{J^{\frac{2(\mu+1)}{3}}},
\end{equation*}
we obtain
\begin{eqnarray*}
{\cal E}_{J}^{\mu}(\rho)&=&{\cal E}_{{\alpha_1}^{\frac{\mu}{\mu+1}}J}^{\mu}(\rho_1)+{\cal E}_{{\alpha_2}^{\frac{\mu}{\mu+1}}J}^{\mu}(\rho_2)-\int\!\!\!\int\frac{\rho_1(x)\rho_2(y)}{|x-y|}\,dxdy\\
&\geq&R_{m,{\alpha_1}^{\frac{\mu}{\mu+1}}J}^{\mu}(\rho_1)+R_{M-m,{\alpha_2}^{\frac{\mu}{\mu+1}}J}^{\mu}(\rho_2)-\frac{m(M-m)}{R'}\\
&=&M^{\frac{7-2\mu}{3}}J^{\frac{2(\mu+1)}{3}}R_{1,1}^{\mu}\Big({\alpha_1}^{\frac{2\mu}{3}}\big(\frac{m}{M}\big)^{\frac{7-2\mu}{3}}+{\alpha_2}^{\frac{2\mu}{3}}\big(\frac{M-m}{M}\big)^{\frac{7-2\mu}{3}}\Big)\\
&&-\frac{m(M-m)}{R'}\\
&=&R_{M,J}^{\mu}\Big(((\frac{m}{M})^{\frac{7}{3}})^{\frac{7-2\mu}{7}}({\alpha_1}^{\frac{7}{3}})^{\frac{2\mu}{7}}+\big((\frac{M-m}{M})^{\frac{7}{3}}\big)^{\frac{7-2\mu}{7}}\big({\alpha_2}^{\frac{7}{3}}\big)^{\frac{2\mu}{7}}\Big)\\
&&-\frac{m(M-m)}{R'}\\
&\geq&R_{M,J}^{\mu}\big[\alpha_1^{\frac{7}{3}}+\alpha_2^{\frac{7}{3}}\big]^{\frac{2\mu}{7}}\big[(\frac{m}{M})^{\frac{7}{3}}+(\frac{M-m}{M})^{\frac{7}{3}}\big]^{\frac{7-2\mu}{7}}-\frac{m(M-m)}{R'}\\
&\geq&R_{M,J}^{\mu}\big[\alpha_1+\alpha_2\big]^{\frac{2\mu}{7}}\big[(\frac{m}{M})^{\frac{7}{3}}+(\frac{M-m}{M})^{\frac{7}{3}}\big]^{\frac{7-2\mu}{7}}-\frac{m(M-m)}{R'}\\
&=&R_{M,J}^{\mu}\big[(\frac{m}{M})^{\frac{7}{3}}+(\frac{M-m}{M})^{\frac{7}{3}}\big]^{\frac{7-2\mu}{7}}-\frac{m(M-m)}{R'}\\
&\geq&R_{M,J}^{\mu}\big[1-\frac{7}{3}\frac{M-m}{M}\frac{m}{M}\big]^{\frac{7-2\mu}{7}}-\frac{m(M-m)}{R'}\\
\end{eqnarray*}
where we used the scaling relations from Lemma \ref{lm:scaling}, the fact that $R_{M,J}^{\mu}$ is negative, $\alpha_1+\alpha_2=1$, Newton's theorem for spherically symmetric potentials (cf.~\cite{LiLo}), the discrete Hölder's inequality, and the fact that for $x\in[0,1]$ we have
\begin{equation*}
x^{\frac{7}{3}}+(1-x)^{\frac{7}{3}}\leq 1-\frac{7}{3}x(1-x).
\end{equation*}
For any $a$, $b>0$ and $0<\alpha<1$ the following inequality (\cite[Thm.~41]{HaLiPo}) holds
\begin{equation*}
b^{\alpha}-a^{\alpha}\geq\alpha b^{\alpha-1}(b-a).
\end{equation*}
By combining it with the previous estimates we obtain:
\begin{eqnarray*}
{\cal E}_{J}^{\mu}(\rho)-R_{M,J}^{\mu}&\geq& -R_{M,J}^{\mu}\Big(1-\big[1-\frac{7}{3}\frac{M-m}{M}\frac{m}{M}\big]^{\frac{7-2\mu}{7}}\Big)-\frac{m(M-m)}{R'}\\
&\geq&-\frac{7-2\mu}{3}\frac{M-m}{M}\frac{m}{M}R_{M,J}^{\mu}-\frac{m(M-m)}{R'}\\
&=&m(M-m)\big[\frac{1}{R_0}-\frac{1}{R'}\big]
\end{eqnarray*}
which proves the first claim of the lemma. The concentration property is now a corollary of the first claim and it is proven by a contradiction argument, in exactly the same way as it was done in the Step $4$, Section $5$ of \cite{Re1}.
\end{proof}

\subsubsection*{Step 4: Removing the symmetry assumption}

Let $(\rho_n)\subset {\cal F}_{M}^{\mu}$ be a minimizing sequence of the functional ${\cal E}_{J}^{\mu}$. Then the the sequence of spherically symmetric rearrangements $(\rho_n^*)$ is also a minimizing sequence. According to Corollary~\ref{co:bitno}, we conclude that $(\rho_n^*)$ is uniformly bounded in $L^{\frac{2\mu+5}{2\mu+3}}(\mathbb R^3)$ and by theorem of Banach-Alaoglu, we conclude that there exists a subsequence of $(\rho_n^*)$, still denoted by $(\rho_n^*)$, such that
\begin{equation*}
\rho_n^*\rightharpoonup\rho'\quad\textrm{weakly in}\quad L^{\frac{2\mu+5}{2\mu+3}}(\mathbb R^3)
\end{equation*}
for some $\rho'\in L^{\frac{2\mu+5}{2\mu+3}}(\mathbb R^3)$. Because of Lemma~\ref{lm:sim} we know that 
\begin{equation}\label{eq:a}
\lim_{n\to\infty}\int_{|x|\geq R_0}\rho_n^*(x)\,dx=0,\quad n\in\mathbb N
\end{equation}
where we choose $R_0$ like in Lemma~\ref{lm:sim}. This fact combined with the weak convergence of $(\rho_n^*)$ easily implies 
\begin{equation*}
\mathrm{supp}(\rho')\subset B_{R_0},\quad\int\rho'\,dx=M.
\end{equation*}
Lemma \ref{lm:novembar} now implies 
\begin{equation*}
\lim_{n\to\infty} {\cal E}_{\textrm{pot}}(\rho_n^*-\rho')=0.
\end{equation*}
By convexity of $\Psi$ and by Mazur's lemma it is easy to deduce that
\begin{equation*}
\Big(\int\Psi(\rho')\,dx\Big)^{\frac{2\mu+3}{3}}\leq\limsup_{n\to\infty}\Big(\int\Psi(\rho_n^*)\,dx\Big)^{\frac{2\mu+3}{3}}, 
\end{equation*}
(cf.~\cite{Ha} or \cite{Re1}). This implies immediately that $\rho'$ is a minimizer and hence
\begin{equation*}
\Big(\int\Psi(\rho_n^*)\,dx\Big)^{\frac{2\mu+3}{3}}\to\Big(\int\Psi(\rho')\,dx\Big)^{\frac{2\mu+3}{3}}.
\end{equation*}
Moreover, 
\begin{eqnarray*}
{\cal E}_{\textrm{pot}}(\rho_n)&=&{\cal E}_{J}^{\mu}(\rho_n)-\Big(\int\Psi(\rho_n)\,dx\Big)^{\frac{2\mu+3}{3}}={\cal E}_{J}^{\mu}(\rho_n)-\Big(\int\Psi(\rho_n^*)\,dx\Big)^{\frac{2\mu+3}{3}}\\
&\to&{\cal E}_{J}^{\mu}(\rho'))-\Big(\int\Psi(\rho')\,dx\Big)^{\frac{2\mu+3}{3}}={\cal E}_{\textrm{pot}}(\rho').
\end{eqnarray*}
We apply now (\cite[Thm.~1]{BuGu}) to conclude that there exists a sequence of translations $T_n'$ such that 
\begin{equation*}
\lim_{n\to\infty}||\nabla U_{T_n'\rho_n}-\nabla U_{\rho'}||_2=0.
\end{equation*}
We easily see that $\lim_{n\to\infty}{\cal E}_{\textrm{pot}}(\rho_n^*)=\lim_{n\to\infty}{\cal E}_{\textrm{pot}}(T_n'\rho_n)={\cal E}_{\textrm{pot}}(\rho')$.
Let us now set $R:=3R_0$. Due to Riesz's rearrangement inequality, following the splitting idea from Lemma~3.1 in \cite{BuGu} (Confinement to a ball), we obtain:
\begin{eqnarray*}
2{\cal E}_{\textrm{pot}}(\rho_n^*)-2{\cal E}_{\textrm{pot}}(\rho_n)&\geq&\int\int{\rho_n^*(x)\rho_n^*(y)\min\big[\frac{1}{|x-y|},\frac{1}{2R_0}\big]}\,dxdy\\
&-&\int\int{\rho_n(x)\rho_n(y)\min\big[\frac{1}{|x-y|},\frac{1}{2R_0}\big]}\,dxdy\geq0.
\end{eqnarray*}
By adding and subtracting the quantity
\begin{equation*}
\int\int\rho_n^*(x)\rho_n^*(y)\frac{1}{2R_0}\,dxdy
\end{equation*}
we get the following:

\begin{eqnarray*}
\lefteqn{2{\cal E}_{\textrm{pot}}(\rho_n^*)-2{\cal E}_{\textrm{pot}}(\rho_n)\geq\int\int\rho_n^*(x)\rho_n^*(y)\frac{1}{2R_0}\,dxdy}\\
&&\mbox{}-\int\int{\rho_n(x)\rho_n(y)\min\big[\frac{1}{|x-y|},\frac{1}{2R_0}\big]}\,dxdy\\
&&\mbox{}+\int\int{\rho_n^*(x)\rho_n^*(y)\min\big[\frac{1}{|x-y|},\frac{1}{2R_0}\big]\,dxdy}-\int\int\rho_n^*(x)\rho_n^*(y)\frac{1}{2R_0}\,dxdy\\
&&=\int\int\rho_n(x)\rho_n(y)\Big[\frac{1}{2R_0}-\min\big[\frac{1}{|x-y|},\frac{1}{2R_0}\big]\Big]\,dxdy\\
&&\mbox{}+\int\int\rho_n^*(x)\rho_n^*(y)\Big[\min\big[\frac{1}{|x-y|},\frac{1}{2R_0}\big]-\frac{1}{2R_0}\Big]\,dxdy\\
&&=\int\int\rho_n(x)\rho_n(y)\Big[\frac{1}{2R_0}-\min\big[\frac{1}{|x-y|},\frac{1}{2R_0}\big]\Big]\,dxdy\\
&&\mbox{}+\int\int_{|x|\geq R_0\vee|y|\geq R_0}\rho_n^*(x)\rho_n^*(y)\Big[\min\big[\frac{1}{|x-y|},\frac{1}{2R_0}\big]-\frac{1}{2R_0}\Big]\,dxdy\\
&&\geq\Big[\frac{1}{2R_0}-\frac{1}{R}\Big]\int\int_{|x-y|\geq R}{\rho_n(x)\rho_n(y)}\,dxdy\\
&&\mbox{}+\int\int_{(|x|\geq R_0\vee|y|\geq R_0)\wedge(|x-y|\geq 2R_0)}{\frac{\rho_n^*(x)\rho_n^*(y)}{|x-y|}}\,dxdy\\
&&\mbox{}-\frac{1}{2R_0}\int\int_{(|x|\geq R_0\vee|y|\geq R_0)\wedge(|x-y|\geq 2R_0)}{\rho_n^*(x)\rho_n^*(y)}\,dxdy\\
&&=: A_n+B_n-C_n.
\end{eqnarray*}
Here we used the equimeasurability of the rearrangements. 
According to the proof of Lemma~3.1 in \cite{BuGu} we conclude that there exists a translation $T_n$ such that $A_n\geq\Big[\frac{1}{2R_0}-\frac{1}{R}\Big]\Big(\int\int_{|x|\geq R}T_n\rho_n(x)\,dx\Big)^2$. By letting $n\to\infty$ it is a direct consequence of (\ref{eq:a}) that both $B_n$ and $C_n$ tend to $0$ as $n\to\infty$. So we get 
\begin{equation}\label{eq:b}
\int_{|x|\geq R} T_n\rho_n(x)\,dx\to0\quad\textrm{as}\quad n\to\infty.
\end{equation}
Since $T_n\rho_n$ is a minimizing sequence, it is uniformly bounded in $L^{\frac{2\mu+5}{2\mu+3}}(\mathbb R^3)$ which implies that there exists some $\rho_0$ such that $T_n\rho_n\rightharpoonup\rho_0$ weakly in $L^{\frac{2\mu+5}{2\mu+3}}(\mathbb R^3)$, and  (\ref{eq:b}) implies
\begin{equation*}
\int_{B_R}{\rho_0}\,dx=M\quad\textrm{and}\quad\mathrm{supp}(\rho_0)\subset B_R.
\end{equation*}
Now by Lemma~\ref{lm:novembar} we conclude   
\begin{equation}\label{eq:sestra}
\nabla U_{T_n\rho_n}\to\nabla U_{\rho_0}\quad\textrm{strongly in}\quad L^2(\mathbb R^3),\quad n\to\infty,
\end{equation}
which, again combined with the convexity of the functional $\Psi(\rho)$, allows for the conclusion that $\rho_0$ is a minimizer of our variational problem. Eqn.~(\ref{eq:sestra}) also implies 
\begin{equation*}
\lim_{n\to\infty}\int (T_n\rho_n)^{\frac{2\mu+5}{2\mu+3}}\,dx=\int\rho_0^{\frac{2\mu+5}{2\mu+3}}\,dx,
\end{equation*}
which means that $||T_n\rho_n||_{L^{\frac{2\mu+5}{2\mu+3}}(\mathbb R^3)}$ converges to $||\rho_0||_{L^{\frac{2\mu+5}{2\mu+3}}(\mathbb R^3)}$ and this fact, combined with the weak convergence, implies the strong convergence in the space $L^{\frac{2\mu+5}{2\mu+3}}(\mathbb R^3)$. This completes the proof of Theorem~\ref{th:c}.  

\textbf{Acknowledgement:} I would like to thank Gerhard Rein for introducing me to the problem and for many stimulating discussions.

\end{document}